\documentclass[12pt]{article}
\usepackage{times}
\usepackage{geometry}
\usepackage[utf8]{inputenc}
\usepackage{enumitem,amssymb}
\usepackage{ragged2e}
\newlist{thematic}{itemize}{8}
\setlist[thematic]{label=$\square$}
\usepackage{pifont}
\newcommand{\cmark}{\ding{51}}%
\newcommand{\done}{\rlap{$\square$}{\raisebox{2pt}{\large\hspace{1pt}\cmark}}%
\hspace{-2.5pt}}

\usepackage{natbib}
\usepackage[colorinlistoftodos]{todonotes}
\usepackage{aas_macros}
\usepackage{graphicx}
\usepackage{mathabx}
\usepackage{makecell}
\def\micron{$\mu$m}

\usepackage{hyperref} 

\begin{document}
\pagestyle{empty}
\raggedright
\huge
Astro2020 Science White Paper \linebreak

Directly Imaging Rocky Planets from the Ground\linebreak
\normalsize

\noindent \textbf{Thematic Areas:} \hspace*{60pt} \done~Planetary Systems \hspace*{10pt} $\square$ Star and Planet Formation \hspace*{20pt}\linebreak
$\square$ Formation and Evolution of Compact Objects \hspace*{31pt} $\square$ Cosmology and Fundamental Physics \linebreak
  $\square$  Stars and Stellar Evolution \hspace*{1pt} $\square$ Resolved Stellar Populations and their Environments \hspace*{40pt} \linebreak
  $\square$    Galaxy Evolution   \hspace*{45pt} $\square$             Multi-Messenger Astronomy and Astrophysics \hspace*{65pt} \linebreak
  
\textbf{Principal Author:}

Name: B. Mazin	
 \linebreak						
Institution: University of California Santa Barbara  
 \linebreak
Email: bmazin@ucsb.edu
 \linebreak
Phone: (805)893-3344
 \linebreak
 
\textbf{Co-authors:} (names and institutions)
  \linebreak
\'{E}. Artigau, Université de Montréal\\
V. Bailey, California Institute of Technology/JPL \\
C. Baranec, University of Hawaii \\
C. Beichman. California Institute of Technology/JPL \\
B. Benneke, Université de Montréal\\
J. Birkby, University of Amsterdam\\
T. Brandt, University of California Santa Barbara \\
J. Chilcote, University of Notre Dame \\
M. Chun, University of Hawaii\\
L. Close, University of Arizona\\
T. Currie, NASA-Ames Research Center\\
I. Crossfield, Massachusetts Institute of Technology \\
R. Dekany, California Institute of Technology \\
J.R. Delorme, California Institute of Technology/JPL \\
C. Dong, Princeton University \\
R. Dong, University of Victoria \\
R. Doyon, Université de Montréal \\
C. Dressing, University of California Berkeley \\
M. Fitzgerald, University of California Los Angeles \\
J. Fortney, University of California Santa Cruz \\
R. Frazin, University of Michigan \\
E. Gaidos, University of Hawai`i \\
O. Guyon, University of Arizona/Subaru Telescope \\
J. Hashimoto, Astrobiology Center of NINS \\
L. Hillenbrand, California Institute of Technology \\
A. Howard, California Institute of Technology  \\
R. Jensen-Clem, University of California Berkeley \\
N. Jovanovic, California Institute of Technology \\
T. Kotani, Astrobiology Center of NINS \\
H. Kawahara, University of Tokyo \\
Q. Konopacky, University of California San Diego\\
H. Knutson, , California Institute of Technology \\
M. Liu, University of Hawaii \\
J. Lu, University of California Berkeley\\
J. Lozi, Subaru Telescope \\
B. Macintosh, Stanford University \\
J. Males, University of Arizona \\
M. Marley, NASA Ames \\
C. Marois, University of Victoria \\
D. Mawet, California Institute of Technology/JPL \\
S. Meeker, California Institute of Technology/JPL \\
M. Millar-Blanchaer, California Institute of Technology/JPL \\
S. Mondal, S. N. Bose National Centre for Basic Sciences \\
N. Murakami, Hokkaido University \\
R. Murray-Clay University of California Santa Cruz \\
N. Narita, National Astronomical Observatory of Japan \\
T.S. Pyo, National Astronomical Observatory of Japan \\
L. Roberts, California Institute of Technology/JPL  \\
G. Ruane, California Institute of Technology/JPL  \\
G. Serabyn, California Institute of Technology/JPL  \\
A. Shields, University of California Irvine \\
A. Skemer, University of California Santa Cruz \\
L. Simard, Herzberg Astronomy and Astrophysics Research Centre \\
D. Stelter, University of California Santa Cruz \\
M. Tamura, University of Tokyo, Astrobiology Center of NINS \\
M. Troy, California Institute of Technology/JPL \\ 
G. Vasisht, California Institute of Technology/JPL \\
J. K. Wallace, California Institute of Technology/JPL \\ 
J. Wang, The Ohio State University \\
J. Wang, California Institute of Technology \\
S. Wright, University of California San Diego \\


\pagebreak
\pagestyle{plain}
\setcounter{page}{1}
\justifying
\section{Introduction}
Over the past three decades instruments on the ground and in space have discovered thousands of planets outside the solar system. These observations have given rise to an astonishingly detailed picture of the demographics of short-period planets ($P\lesssim 30$ days), but are incomplete at longer periods where both the sensitivity of transit surveys and radial velocity signals plummet. Even more glaring is that the spectra of planets discovered with these indirect methods are either inaccessible (radial velocity detections) or only available for a small subclass of transiting planets with thick, clear atmospheres~\citep{2015ApJ...815..110M}.

Direct detection can be used to discover and characterize the atmospheres of planets at intermediate and wide separations, including non-transiting exoplanets. Today, a small number of exoplanets have been directly imaged, but they represent only a rare class of young, self-luminous super-Jovian-mass objects orbiting tens to hundreds of AU from their host stars. Atmospheric characterization of planets in the $<$5 AU regime, where radial velocity (RV) surveys have revealed an abundance of other worlds, is technically feasible with 30-m class apertures in combination with an advanced AO system, coronagraph, and suite of spectrometers and imagers. 

There is a vast range of unexplored science accessible through astrometry, photometry, and spectroscopy of rocky planets, ice giants, and gas giants. In this whitepaper we will focus on one of the most ambitious science goals --- detecting for the first time habitable-zone rocky ($<1.6 R_\Earth$, \citealt{2017AJ....154..109F}) exoplanets in reflected light around nearby M-dwarfs\footnote{cf. whitepaper to the Exoplanet Science Strategy committee of the U.S. National Academy of Science: \url{https://goo.gl/nELRGX}}.  Other whitepapers will address the second potential way to detect rocky exoplanets by looking at thermal (10~\micron) emission around a handful of the nearest Sun-like stars.  

To truly understand exoplanets we need to go beyond detection and demographics and begin to characterize these exoplanet and their atmospheres.  High-resolution spectroscopic capabilities will not only illuminate the physics and chemistry of exo-atmospheres, but may also probe these rocky, temperate worlds for signs of life in the form of atmospheric biomarkers (combination of water, oxygen, methane and other molecular species). The Astro 2020 White Paper "Detecting Earth-like Biosignatures on Rocky Exoplanets around Nearby Stars with Ground-based Extremely Large Telescopes" (Lopez-Morales et al.) includes more details on potential biosignatures. 

\vspace{-0.2in}
\section{Landscape and Demographics}
Led by \textit{Kepler}, transit surveys of the last decade have revealed that about half of all solar-type stars host planets with sizes between
Earth and Neptune (1--4 $R_\Earth$) and orbital periods less than a year, and that these planets are frequently found in closely packed multiple
systems. The occurrence rate of such small planets declines at the shortest orbital periods ($P<$10\,days), but is roughly flat in $\log P$ for
periods between a month and a year~\citep[ and refs. therein]{winn&fabrycky15}. Jovian worlds are known to be less common than terrestrial
planets~\citep{howard_etal10, mayor_etal11}, but their higher detectability means that they comprise the bulk of the presently detected population
of long-period planets ($P > 1$\,year; NASA Exoplanet Archive). At the longest periods, current imaging surveys of young systems have found that
very massive planets are rare (0.6\% for 5--13 $M_\textrm{Jup}$ and $a = $30--300\,AU;~\citealt{bowler_etal16}), but true Jupiter analogs are beyond
the reach of existing instruments. Even with instrumentation that reaches fundamental limits (AO correction limited by photon shot noise, near
perfect coronagraphs, etc.) the science return of 8--10\,m telescopes is fundamentally limited by inner working angle (IWA) and sensitivity, but we
do expect future surveys on existing telescopes to reveal both cooler and less massive (perhaps sub-Jovian at the youngest ages) planets at
moderate separations from their host stars.

While \textit{Kepler} has probed the landscape and demographics of small exoplanets on short period orbits, and RV surveys have given valuable information for wider orbits, the census of small planets in the solar neighborhood is highly incomplete.  \textit{Kepler} has also not provided information on small planets with intermediate periods (above 200 days).  A large ground-based telescope, like those being considered in the US ELT program, could fill in this important gap in our knowledge, potentially in conjunction with upcoming high precision RV surveys.

The large increase in aperture going from current telescopes to 30-m class telescopes both decreases the inner working angle, probes smaller separations, and increases the achievable contrast (even with equal wavefront control performance, as the fractional stellar flux per $\lambda/D$ element scales as $1/D^2$).  As our understanding of the challenges of direct imaging and our technology continues to improve, we expect direct imaging instruments for ELTs to approach photon-noise limited final post-processed contrasts of around $10^{-8}$~\citep{males_guyon_2018,2018SPIE10703E..0ZG}.

The IWA of 30-m telescopes will enable access to the habitable zones of hundreds of cool dwarfs.  This synergizes well with future NASA missions like HabEx and LUVOIR that plan to search for earth-like planets around Sun-like stars (see also Astro2020 White Paper "The Critical, Strategic Importance of Adaptive Optics-Assisted Ground-Based Telescopes for the Success of Future NASA Exoplanet Direct Imaging Missions" by Currie et al.).  Together, ground-based 30-m telescopes working in the visible and NIR and future visible space-based planet finders will provide a complete picture of habitable zone planets in the solar neighborhood.

Outside M-dwarf habitable zones, these large ground-based telescopes will also have the ability to greatly contribute to the demographics of planets from 0.5 to beyond 5\,AU, especially around stellar types that are not amenable to RV measurements and for face-on systems. 

\begin{figure*}\centering 
\vspace{-.7in}
\includegraphics[width=0.47\textwidth]{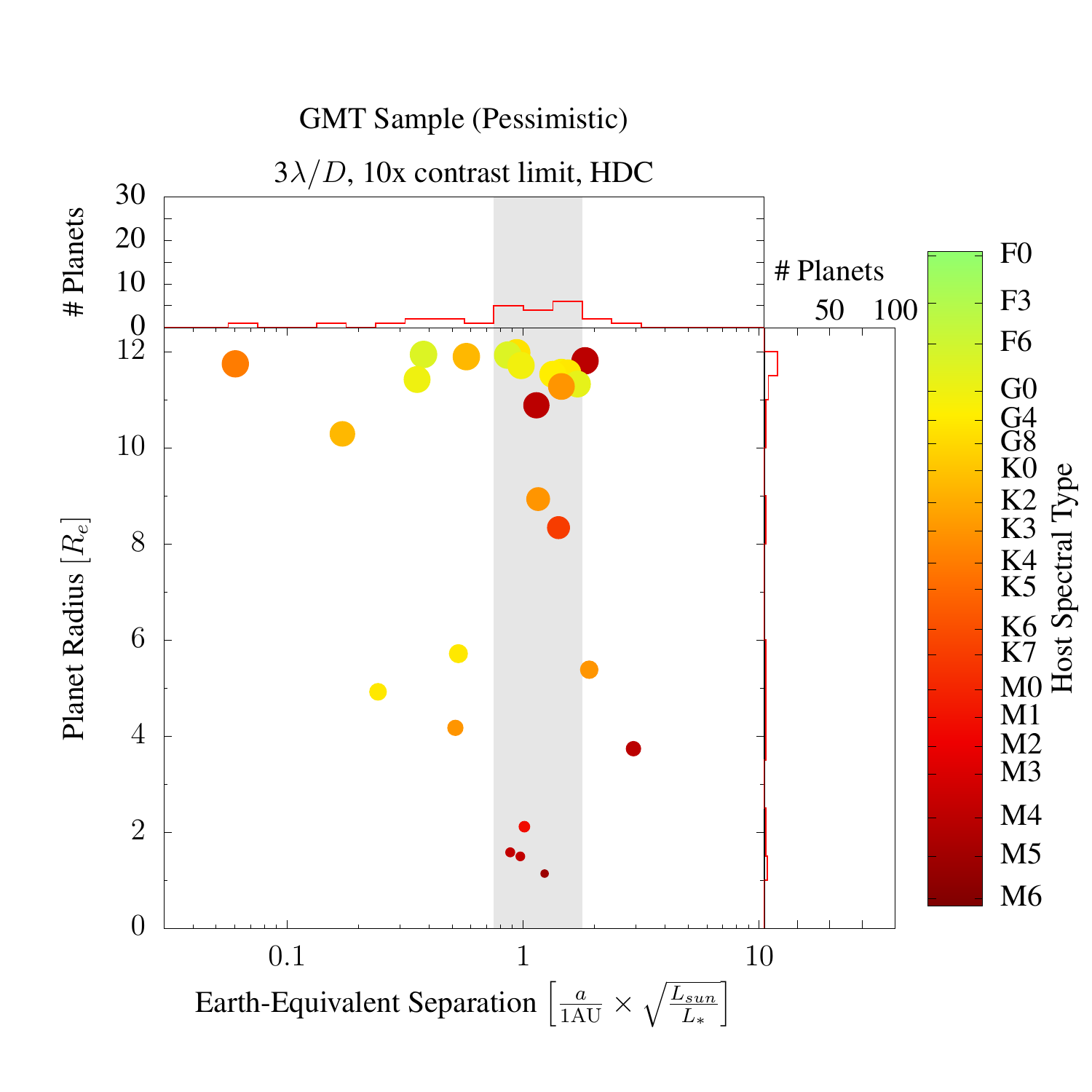}
\includegraphics[width=0.47\textwidth]{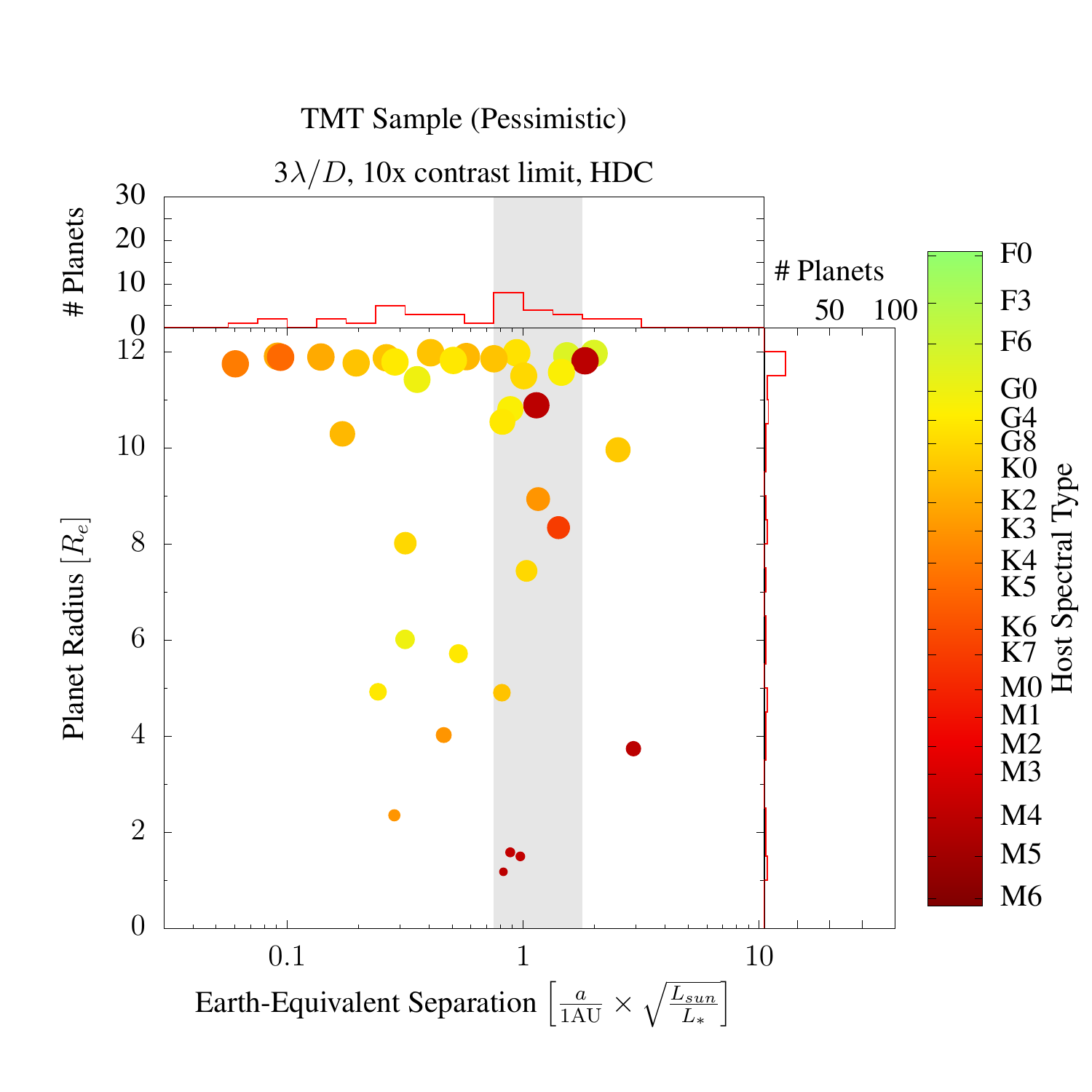}
\includegraphics[width=0.47\textwidth]{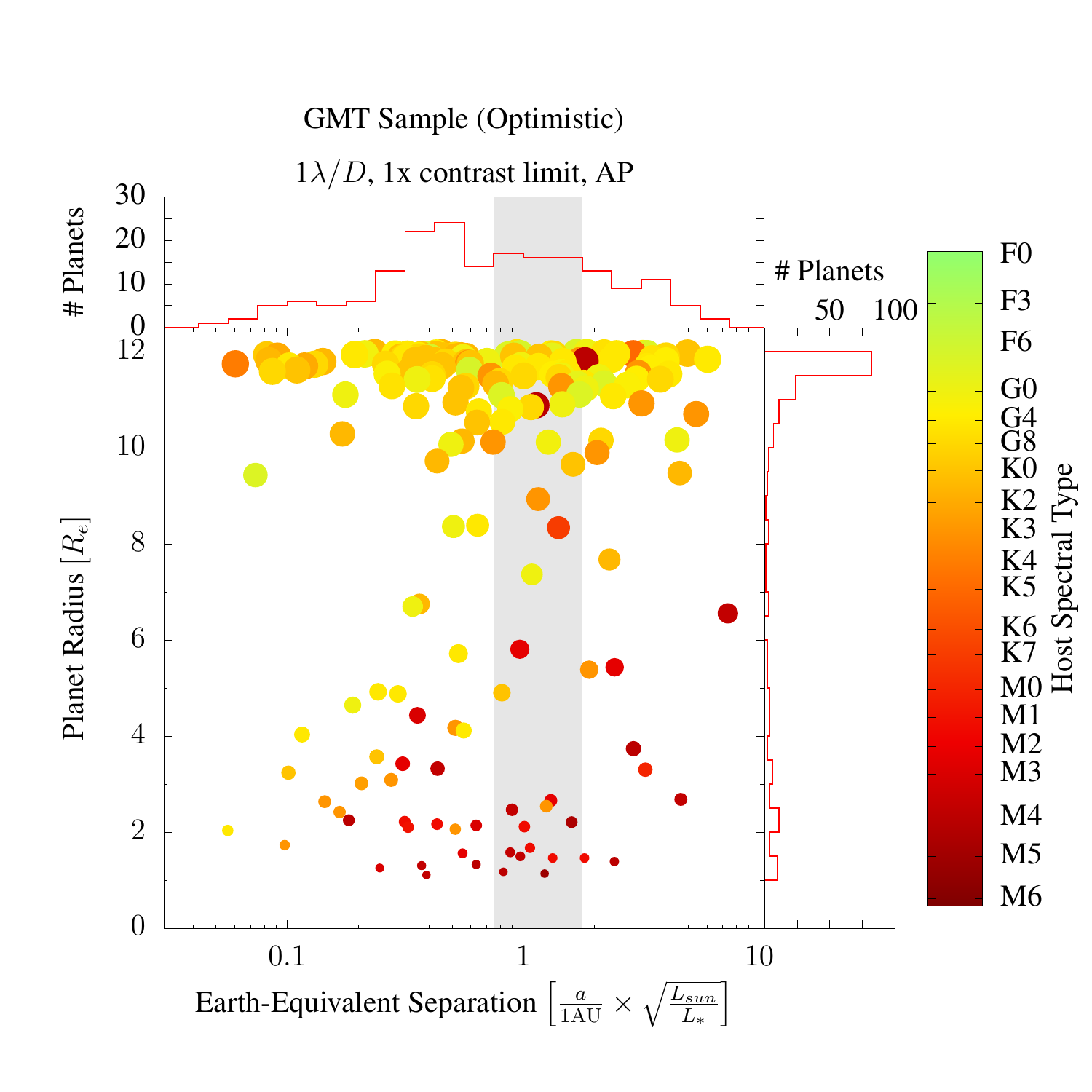}
\includegraphics[width=0.47\textwidth]{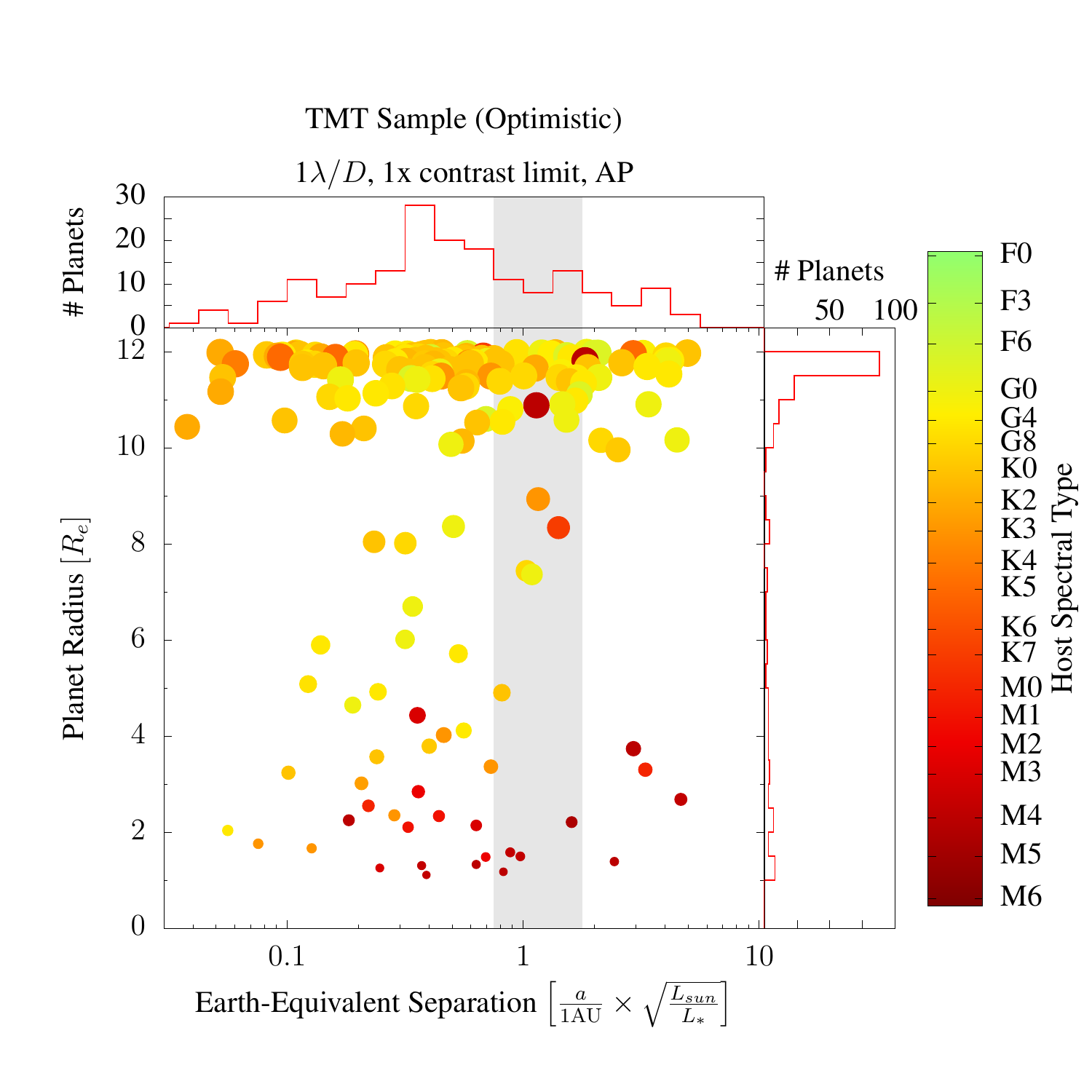}
\vspace{-.3in}
\caption{\small Surveying a significant sample of both giant ($R_{Jup} = 11 R_{\Earth}$) and rocky ($<1.6 R_{\Earth}$) planets across a range of equilibrium temperatures will probe regions where condensates like H$_2$O, NH$_3$, and CH$_4$ are expected to play a major role in regulating planet formation. The above plots show we can achieve these science goals with a sample of \textbf{currently} known planets (primarily detected through radial velocities) with GMT (left panels) and TMT (right panels) in a survey spanning only 28 nights of integration.  The shaded region denotes an estimate of the habitable zone. GMT actually shows more small planet detections despite the smaller aperture because the RV coverage in the Southern Hemisphere has historically been better (HARPS). The top panels assume the pessimistic case of a $3\lambda/D$ IWA coronagraph, and a residual stellar halo of 10$\times$ the limits of \citet{males_guyon_2018}. The residual speckles are pessimistically assumed to be long lived, so the high-dispersion coronagraphy (HDC) technique is used to cross-correlate the reflected stellar spectrum.  Spectral-type specific templates were used to derive the cross-correlation signal.  Even with degraded instrument performance, several dozen known planets will be characterized in reflected light.  The bottom panels assume a 1 $\lambda/D$ IWA, photon noise, and short-lived residual speckle noise after real-time control of quasi-static aberrations and predictive control of the atmosphere.  Now aperture photometry is more efficient than HDC.  An empirical planetary mass-radius relationship is used to derive planet radii.   \textbf{These plots bound the potential of TMT and GMT for reflected-light characterization of exoplanets, and should motivate significant efforts toward optimizing ground-based instruments for direct imaging.  As this plot is only for currently known planets it is not an exhaustive target list --- with continued effort, many more planets will be known by ELT first light, giving many more targets without the need for blind surveys.}}
\label{fig:detection_known}
\vspace{-.2in}
\end{figure*}

\vspace{-0.2in}
\section{Planetary Characteristics}
Even more crucial than detecting planets with a variety of masses and separations is probing their atmospheric properties with spectroscopy and their bulk properties through combined mass and radius measurements. The classic ``core accretion'' model of planet formation is supported by the enhanced metallicities and detailed compositions of the giant planets in our solar system and their correlation with planetary mass and semimajor axis. When spectroscopy is combined with atmospheric models it allows us to infer the atmospheric compositions of exoplanets. Ongoing studies of giant exoplanet metallicity are beginning to explore whether core accretion is the dominant formation pathway for most exoplanetary systems. To date, however, observations of water or carbon abundance have been practical only for transiting hot-Jupiter planets and for self-luminous young giant planets discovered through direct imaging. 

Over the next decade, the spectroscopic exploration of transiting planets will advance rapidly with the combination of \textit{TESS} and \textit{JWST}, including a potential sample of $\sim$100 giant planets and many more small planets (albeit biased towards planets with high equilibrium temperatures and/or planets orbiting low-mass stars;~\citealt{sullivan_etal15}). ELTs, however, will measure cooler planets and those orbiting earlier-type stars.  By probing different regimes of atmospheric chemistry than transit observations, we increase the parameter space spanned by our physical models, helping to identify their biases. Together, transit and direct methods will provide the spectroscopy of the diverse array of planetary targets that are needed to significantly further our understanding of planet formation.

In addition to composition, measurements at high spectral resolution can determine planets' rotation rates and cloud through Doppler imaging~\citep{crossfield14,2014Natur.509...63S}.  Photometry and polarimetry as a function of orbital phase can also be used to constrain clouds, hazes, and surface features~\citep[e.g.][]{2008ApJ...676.1319P,karalidi&stam12}.  These science cases are enabled by these telescope's unprecedented angular resolution, light-collecting power, and instrumentation.  Moreover, these instruments have the potential opportunity to detect biosignatures using a technique called high-dispersion coronagraphy~\citep[HDC;][]{2013Sci...339.1398K,2015A&A...576A..59S,2015ApJ...804...61B,wang_etal17,2018A&A...617A.144H}. HDC takes advantage of high-resolution spectroscopy ($R>$50k) to mitigate previous contrast limits from the ground and in principle reach the sensitivities needed to characterize planets that are ${\sim}10^8$ times fainter than their host stars. 

\vspace{-0.2in}
\subsection{Rocky Planets}

Characterising the atmosphere of planets in the habitable zone of nearby stars to search for biosignature gases is perhaps one of the most compelling science goals in all of astronomy.  This importance is reflected, for example, in Strategic Objective 1.1 of the 2018 NASA Strategic Plan which states: \emph{``Are we alone?'' is a central research question that involves biological research and research in the habitability of locations in the solar system such as Mars, the moons of outer planets, or thousands of potentially habitable worlds around other stars. This research supports a fundamental science topic at the interface of physics, chemistry, and biology.}

Around the nearest M dwarfs, ELTs will be able to detect starlight reflected by rocky, habitable-zone exoplanets in addition to ice and gas giants. The low luminosities of M dwarfs means that planets must orbit close to the star to receive Earth-like radiance levels. While a boon for transit surveys due to the increased probability of transit, the proximity of M dwarf habitable zones to the star poses a challenge for direct imaging because the habitable zone typically lies well inside the IWA of 8-m telescopes.  The small IWA of ELTs makes them ideal for characterizing planets in the habitable zones of M dwarfs.  Only a small fraction of nearby planets will transit, and we await an understanding of the success rates of NIR-optimized RV surveys in discovering low-mass planets.  Despite the uncertainties, Proxima Cen b, the TRAPPIST-1 planets, LHS 1140 b, and Gliese 411 (see Table~\ref{table:planets}) were all discovered in the last two years, and several more surveys targeting M dwarfs have just started or will begin in the near future. ELTs will be powerful instruments not only in obtaining more complete and less biased statistics on planetary demographics through surveys that image planets orbiting low-mass stars, but also in characterizing these discoveries.

\begin{table}[htb!]
	\centering
	\small
	\begin{tabular}{llllllll}
		\hline
		\textbf{Name} & \textbf{Telescope} & \textbf{Primary} & \textbf{Dist (pc)} & \textbf{Radius} & \textbf{Temp (K)} & \textbf{Sep $\lambda/D$ (mas)} & \textbf{Contrast} \\
		\hline
		 Prox. Cen b & GMT  & M5.5 & 1.3 & 1.1 R$_\Earth$ & 235 & 4.5 (37) & $3.5 \times 10^{-8}$ \\
		 Gliese 411 & TMT & M1.9 & 2.5 & 1.5 R$_\Earth$ & 350 & 4.5 (31) & $2.5 \times 10^{-8}$ \\
		 Ross 128 b & Both & M4 & 3.4 & 1.1 R$_\Earth $ & 300 & 2.2 (15) & $3.3 \times 10^{-8}$ \\
		 YZ Cet d & Both & M4.5 & 3.6 & 1.1 R$_\Earth $ & 260--370 & 1.1 (7.7) & $9.8 \times 10^{-8}$ \\
		 GJ 273 b & Both & M3.5 & 3.8 & 1.5 R$_\Earth $ & 260 & 3.5 (24) & $2.1 \times 10^{-8}$ \\
		 Wolf 1061 c & Both & M3.5 & 4.3 & 1.7 R$_\Earth $ & 225 & 3.0 (21) & $2.4 \times 10^{-8}$ \\
        \hline
	\end{tabular}
	\caption{The most promising currently known rocky exoplanet candidates with equilibrium temperatures below 350 K from the NASA Exoplanet Archive.  Contrasts are calculated with an albedo of 0.3, phase correction of 50\%, and with radii of the RV planets estimated from models of rocky exoplanets~\citep{2016ApJ...819..127Z}.  Separation in $\lambda/D$ at 1~\micron~is calculated with the most favorable telescope if both TMT and GMT can see the star.  All of these candidates have separations and contrasts that are likely to be accessible with TMT and GMT. Many more planets will be discovered before ELT first light through near-IR RV surveys.}
	\label{table:planets}
\end{table}

These observations will be among the first opportunities to detect biosignatures in the atmospheres of other worlds.  While 10~$\mu$m observations may allow detection of rocky planets around a small sample of solar-type stars~\citep{2015IJAsB..14..279Q}, planets around faint M-type stars are the most favorable targets for spectroscopic follow up at shorter wavelengths with ELTs and \textit{JWST}~\citep[in the case of transiting planets;][]{kasting_etal14}.  There are abundant lines from biosignature gases, O$_2$, H$_2$O, CH$_4$, and CO$_2$, in the near infrared (i.e., $\sim$1 to 4\,\micron), where the HDC technique is expected to reach optimal performance.  For example, a simultaneous detection of oxygen (O$_2$) and methane (CH$_4$) would be highly suggestive of life~\citep{desmarais02}. It is important to note that the very precise characterization of molecular spectra in the laboratory are essential for these detections and should be pursued in support of ELT success.  A detection of CH$_4$, however, is out of reach for \textit{JWST} given the low concentration of CH4 and the relatively high mean molecular weight / small scale height of an Earth-like atmosphere.  Such measurements with ELTs can potentially provide a strong case for life activities on nearby worlds: there are $>$30 M dwarfs within 5 pc that are observable by TMT and GMT, and there is at least one rocky planet per M dwarf~\citep{dressing&charbonneau15}.  Given the many nascent instruments that will undertake M-dwarf planet surveys in the near term, ELTs will likely have access to a full census of characterizable planets in these nearest systems.

\vspace{-0.2in}
\section{Planetary System Architectures}

Significant advances in understanding planet formation are possible by investigating the correlation of planet properties with location in circumstellar disks.  ELTs will be able to place individual planets into formation context in systems that are still forming.  The smaller IWA will allow them to peer into the inner regions of other systems that are inaccessible to today’s telescopes.  This work is discussed in detail in a companion whitepaper, ``Planet Formation Science with US ELT Direct Imaging'' by Dr. Steph Sallum \emph{et al}. 

\bibliographystyle{apj}
\setlength{\bibsep}{0.0pt}
\bibliography{literature,literature2,mazin3}

\end{document}